# $TbPt_6Al_3$: A rare-earth-based $g$-wave altermagnet with a honeycomb structure

R. Oishi[1*†], T. Taniguchi[2], D. T. Adroja[3,4], M. D. Le[3], M. Aouane[3,5], T. Onimaru[1], K. Umeo[6], I. Ishii[1], and T. Takabatake[1]

[1] *Department of Quantum Matter, Graduate School of Advanced Science and Engineering, Hiroshima University, Higashi-Hiroshima 739-8530, Japan,*
[2] *Institute for Materials Research, Tohoku University, Katahira, Sendai 980-8577, Japan,*
[3] *ISIS Facility, STFC, Rutherford Appleton Laboratory, Chilton, Didcot, Oxfordshire OX11 0QX, United Kingdom,*
[4] *Highly Correlated Matter Research Group, Physics Department, University of Johannesburg, Auckland Park 2006, South Africa,*
[5] *European Spallaition Source ESS ERIC, P.O. Box 176, SE-221 00 Lund, Sweden,*
[6] *Department of Low Temperature Experiment, Integrated Experimental Support/Research Division, N-BARD, Hiroshima University, Higashi-Hiroshima 739-8526, Japan*

(Dated: August 22, 2025)

The magnetic properties of the Tb-honeycomb lattice compound $TbPt_6Al_3$, which crystallizes in the $NdPt_6Al_3$-type trigonal structure, have been studied by the measurements of electrical resistivity, magnetization $M(T, B)$, and specific heat on single-crystalline samples. The magnetic susceptibility, $M(T)/B$, for $B \parallel c = 0.1$ T shows a cusp at $T_N = 3.5$ K, which temperature decreases with increasing the magnitude of $B \parallel c$, while $M(T)/B$ for $B \parallel a = 0.1$ T remains constant with decreasing temperature below $T_N$. This anisotropic behavior suggests a collinear antiferromagnetic (AFM) order of the $Tb^{3+}$ moments pointing along the $c$ axis. The data of $M(T)/B$ for $T > 10$ K on the single crystal and that of inelastic neutron scattering from powdered samples have been simultaneously analyzed using the crystal field model. The analysis reveals the non-Kramers doublet ground state for the $Tb^{3+}$ ion under the trigonal crystal field. The neutron powder diffraction measurement shows that the collinear AFM structure with a magnetic propagation vector $\mathbf{k} = [0, 0, 0]$ is associated with moments of 5.1 $\mu_B$/Tb pointing along the $c$ axis. Comparison of the magnetic point group with the nontrivial spin Laue group indicates that $TbPt_6Al_3$ is classified into bulk $g$-wave altermagnets.

*oishi@es.hokudai.ac.jp,
†Present address: *Research Institute for Electronic Science, Hokkaido University, Sapporo 001-0020, Japan*

# I. INTRODUCTION

Time-reversal symmetry (TRS) breaking in the point group is a key ingredient of an anomalous Hall effect, spin-splitting, and piezomagnetic effect [1-6]. Such phenomena have been investigated in ferromagnets, in which all magnetic moments are pointing along the same direction [7]. For the conventional collinear antiferromagnets (AFMs), TRS is preserved since the magnetic ions have a symmorphic symmetry combining a translation and an inversion. However, if the magnetic ions in an equivalent position are connected by non-symmorphic operations such as a screw or glide symmetry (mirror and half translation), TRS can be broken even in a collinear AFM ordered state [2,8,9]. This type of collinear AFM is called altermagnet [10,11] or characterized as a magnetic toroidal quadrupole ordering [12-14]. The altermagnets are categorized by $d$-, $g$-, and $i$-wave symmetry of the spin-splitting band structures. The TRS breaking phenomena in the altermagnets have been investigated mostly for transition-metal-based compounds [15-22], while those for rare-earth-based compounds have been rarely investigated so far [23].

As a candidate of the collinear AFM breaking the TRS, we focus on $R$Pt$_6$Al$_3$ ($R$ = Nd, Sm, Gd, and Tb), which crystallizes in the NdPt$_6$Al$_3$-type trigonal structure with the centrosymmetric space group $R\bar{3}c$ [24]. As shown in Fig. 1(a), the $R$ atoms at the Wyckoff site $12c$ are arranged on the honeycomb structure centered by a triangle of nonmagnetic Pt atoms. The $R$-based honeycomb layer and Pt- and Al-based blocks are stacked alternately along the $c$ axis. Figure 1(b) shows that the $R$ atoms are connected by $c$-glide and have two-fold rotational symmetries without translation and inversion ones. Among the $R$Pt$_6$Al$_3$ series, NdPt$_6$Al$_3$ and SmPt$_6$Al$_3$ order into an AFM structure with a magnetic propagation vector **k** = [0, 0, 0] [25,26]. NdPt$_6$Al$_3$ orders below 1.2 K into a canted AFM state with the Nd moments lying in the honeycomb plane, whose magnetic point group is 2/$m$.1 [25]. SmPt$_6$Al$_3$, on the other hand, orders below 6.5 K into a collinear AFM structure with the Sm moments pointing toward the $c$ axis [26]. A study of a single-crystal resonant x-ray scattering for SmPt$_6$Al$_3$ has selected two plausible models for the magnetic structure. The neutron diffraction technique, however, has not been applied to distinguish the two models due to the large absorption cross-sections of the Sm nuclei for thermal neutrons.

In order to explore the collinear AFM among the $R$Pt$_6$Al$_3$ series, we turn our attention to TbPt$_6$Al$_3$ in which the Tb$^{3+}$ ion with eight $4f$ electrons has a total angular moment of $J$ = 6. Since the number of $4f$ electrons is even, the Tb$^{3+}$ ion under the trigonal crystal electric field (CEF) in TbPt$_6$Al$_3$ has a non-

Kramers doublet whose two-fold degeneracy is not protected by the TRS. Eustermann *et al.* observed a kink in the magnetic susceptibility of a polycrystalline sample of TbPt$_6$Al$_3$ at 3.6 K and argued that this compound orders antiferromagnetically [24].

In this work, we have synthesized a single-crystalline sample of TbPt$_6$Al$_3$ and studied the magnetic and transport properties by bulk measurements, which confirms the AFM transition at $T_N$ = 3.5 K. The CEF excitations have been measured by the inelastic neutron scattering (INS). At temperatures below $T_N$, magnetic Bragg peaks have been observed by neutron powder diffraction (NPD) measurements. The analysis of the combined results reveals the magnetic anisotropy, CEF effect, and collinear AFM structure. The feasibility of TbPt$_6$Al$_3$ as an altermagnet will be discussed.

## II. Crystal growth and characterizations

The process of single crystal growth of TbPt$_6$Al$_3$ was started with the preparation of a polycrystalline ingot of 15 g. The metallic elements with composition Tb : Pt : Al = 1 : 6 : 3.03 were reacted in a Hukin-type crucible by RF heating [27]. The ingot was melted completely in a tungsten crucible when it was heated up to 1350°C. By pulling a seed rod at a speed of 8 mm/h, we obtained a crystal rod of 50 mm in length as shown in Fig. 1(c). The atomic compositions of several parts were examined by wavelength dispersive electron-probe microanalysis (EPMA), resulting in TbPt$_{5.91}$Al$_{2.86}$. The slight deficit in Pt and Al with respect to TbPt$_6$Al$_3$ is due to the precipitation of an impurity phase of PtAl. The trigonal crystal structure of TbPt$_6$Al$_3$ was confirmed by a powder x-ray diffraction with Cu $K\alpha$ radiation. A Rietveld refinement of the diffraction pattern at room temperature yielded trigonal lattice parameters as $a$ = 7.5471(2) Å and $c$ = 39.252(1) Å, which values agree with the reported ones [24]. For neutron scattering and diffraction experiments, we prepared polycrystalline samples of TbPt$_6$Al$_3$ by arc-melting and subsequent annealing at 1100°C for 10 days.

## III. Experimental Procedures

For the measurements of transport, magnetic, and specific-heat measurements, the single crystal was oriented to the trigonal $a$ and $c$ axes by the back-reflection Laue x-ray pattern. The oriented crystals were cut into the dimensions by spark erosion. A four-terminal ac method was used for the measurement of temperature-dependent resistivity $\rho(T)$ from 0.05 to 3 K in various constant magnetic fields up to 6 T with an adiabatic demagnetization refrigerator mF-ADR50. A Gifford-McMahon-type refrigerator was used for $\rho(T)$ measurements in zero field from 3 to 300 K. We used a Quantum Design MPMS SQUID magnetometer for the measurements of magnetization $M(T, B)$ from 1.8 to 300 K in magnetic fields $B$ up to 5 T. The isothermal magnetization $M(B)$ up to 14 T was measured by the extraction method with a pair detection coils in the insert of ac Measurement System of Quantum Design PPMS. The measurement of the specific heat $C(T)$ from 0.4 to 20 K was performed in constant external fields up to 3 T by the relaxation method on the PPMS.

The INS spectra were collected by using the neutron time-of-flight spectrometer MARI at the ISIS facility in Rutherford Appleton Laboratory [28]. The powdered sample of 10g was wrapped by Al-foil and mounted inside a cylindrical Al-can. Measurements were carried out at 5 K with an incident energy $E_i$ of 8.75 meV [29]. The collected data were analyzed using the software MANTID [30] to determine the CEF level scheme. The NPD experiments were performed with the powder diffractometers HERMES stationed at the Japan Research Reactor [31]. The sample of 10 g was introduced in a cylindrical vanadium sample holder, which was set in a $^3$He cryostat. The NPD data were obtained in zero field at temperatures between 0.7 and 6 K.

## IV. RESULTS AND DISCUSSION

### A. Electrical resistivity

As shown in Fig. 2(a), the electrical resistivity $\rho(T)$ of TbPt$_6$Al$_3$ for $I \parallel a$ and $I \parallel c$ decreases linearly on cooling from 300 to 30 K. Upon further cooling, $\rho(T)$ data bend at around 3.4 K, as better seen in Figs. 2(b) and 2(c). This temperature agrees with the reported $T_N$ of 3.6 K at the kink of magnetic susceptibility for the polycrystalline sample [24]. With increasing $B \parallel a$ to 1 T, the bend in the $\rho(T)$ curve stays at 3.2 K. When $B \parallel c$ is increased to 1 T, on the other hand, $T_N$ gradually decreases to 2.4 K, in consistent with an AFM order. The large values 75 and 56 μΩ cm for $I \parallel a$ and $I \parallel c$, respectively, for the residual resistivity may reflect the atomic disorder in the nonmagnetic sublattice due to the slight deficit in Pt and Al as suggested by EPMA.

### B. Magnetic susceptibility and magnetization

The inverse magnetic susceptibility data of TbPt$_6$Al$_3$ in $B = 0.1$ T are plotted vs $T$ in the inset of Fig. 3. Above 100 K, $B/M(T)$ data are fitted by a Curie-Weiss form, $M(T)/B = N_A\mu_{\text{eff}}^2/3k_B(T - \theta_p)$, where $N_A$ is the Avogadro's number, $\mu_{\text{eff}}$ the effective magnetic moment, and $k_B$ the Boltzmann's constant, $\theta_p$ the paramagnetic Curie temperature. For both $B \parallel a$ and $B \parallel c$, $\mu_{\text{eff}}$ was obtained as 9.9 $\mu_B$/f.u, which is close to 9.72 $\mu_B$ expected for a free Tb$^{3+}$ ion. The negative values of $\theta_{\parallel a}$ = −3.2 K and $\theta_{\parallel c}$ = −24 K indicate the dominant AFM interactions. As shown in the main panel of Fig. 3, $M(T)/B$ for $B \parallel a$ becomes larger than that for $B \parallel c$ on cooling below 100 K. This easy-plane magnetic anisotropy in the paramagnetic state is explained by the trigonal CEF effect on the Tb$^{3+}$ ions as described below.

Upon further cooling, $M(T)/B$ for $B \parallel a$ bends at $T_N$ and becomes flat, as expected for the AFM order in fields applied perpendicular to the direction of the ordered moments [32]. For $B \parallel c$, on the other hand, $M(T)/B$ exhibits a kink and decreases largely, as expected for the AFM order in fields applied parallel to the direction of the ordered moments. As shown in Fig. 4(b), when $B \parallel c$ is increased to 1 T, $T_N$ decreases from 3.3 K to 2.4 K. At $B \parallel c$ = 2 T, no anomaly is observed down to 1.8 K. The large decrease in $M(T)/B_{\parallel c}$ below $T_N$ and the suppression of $T_N$ for $B \parallel c$ indicate that the ordered moments of the Tb$^{3+}$ ions are parallel to the $c$ axis despite being the magnetic hard axis in the paramagnetic state.

Figure 5 represents the isothermal magnetization $M(B)$ at 1.8 K for $B \parallel a$ and $B \parallel c$. In the whole range of $B$ up to 5 T, the value of $M(B)$ for $B \parallel a$ is larger than that for $B \parallel c$, which is consistent with the easy-plane anisotropy in $M(T)/B$ (see Fig. 3). Note that $M(B \parallel c)$ exhibits a metamagnetic increase at around 0.9 T. A linear extrapolation from the data in the range between 1.4 and 2 T goes to the origin $B = 0$, indicating the metamagnetic behavior to be a spin flop transition [32]. On the contrary, $M(B \parallel a)$ continues to increase linearly with $B$ up to 2 T. Upon further increasing $B$ to 5 T, the value of $M(B)$ for $B \parallel a$ and $B \parallel c$ reach 6.5 and 5.6 $\mu_B$/f.u., respectively. Owing to the CEF effect, these values are smaller than 9 $\mu_B$ that is expected for the full polarization of the magnetic moments of a Tb$^{3+}$ free ion.

### C. Specific heat

The specific heat $C(T)$ data give us information about the degeneracy of the ground state and the CEF level scheme. As shown in the inset of Fig. 6(a), $C(T)$ of $R$Pt$_6$Al$_3$ for $R$ = Tb, $C_{Tb}$, jumps at $T_N$ = 3.5 K and displays a lambda-type anomaly, indicating a second-order phase transition. When $B \parallel c$ is increased, $T_N$ gradually decreases and the anomaly disappears at 3 T. By assuming the $C(T)$ data for YPt$_6$Al$_3$ [33], $C_Y$, as the sum of phonon and conduction electron contributions in $C_{Tb}$, we estimate the magnetic contribution as $C_m = C_{Tb} - C_Y$. As shown in Fig. 6 (a), $C_m$ peaks at $T_N$ and has a broad maximum at around 8 K, which is attributed to the Schottky anomaly due to the thermal excitations among CEF levels.

On cooling below 1 K, $C_m(T)$ shows an upswing. Since the temperature dependence of the upswing in $C_{Tb} - C_Y$ is proportional to $T^{-2}$, it is ascribed to a Schottky tail of the nuclear specific heat of Tb$^{159}$ nuclei. We fitted $C_m(T)$ data below 1.5 K with the expression $C_m = \gamma + A_n T^{-2} + \alpha T^n \exp(-\Delta/k_B T)$, where $\gamma$ is a Sommerfeld coefficient, the second term is a nuclear Schottky term, and the last one is the phenomenologically $T$-dependent part of 4$f$ magnetic excitations [34]. The fit gives $\gamma \simeq 0.005$ J K$^{-2}$mol$^{-1}$ and $A_n \simeq 1.6$ J K mol$^{-1}$. This value of $\gamma$ is close to that for the isostructural compound NdPt$_6$Al$_3$ [25]. The specific heat contribution from the 4$f$ electrons $C_{4f} = C_m - A_n T^{-2}$ is plotted in Fig. 6(b) together with the magnetic entropy $S_{4f}(T)$. The $S_{4f}$ data was calculated by integrating the $C_{4f}/T$ data with respect to $T$. The value of $S_{4f}(T)$ at $T_N$ reaches 1.2 $R$ln2, where $R$ is the gas constant. This fact suggests that the CEF ground state of TbPt$_6$Al$_3$ is a non-Kramers doublet. The 20% excess at $T_N$ is attributed to the contribution from the CEF excited levels as manifested itself as the Schottky anomaly in $C_m(T)$ at around 8 K.

We constructed the $B$-$T$ phase diagram as shown in Fig. 7, where the temperature at anomalies in $\rho(T)$, $M(T)$, and $C(T)$ measurements in various constant magnetic fields are plotted. The application of $B \parallel c$ more strongly suppresses the AFM order than that for $B \parallel a$. Especially for $B \parallel c$, the spin-flop transition occurs at around 1 T for $T < 2.5$ K. These results suggest that the Tb$^{3+}$ moments are collinearly oriented along the $c$ axis in the AFM ordered state.

### D. Inelastic neutron scattering

In order to understand the magnetic anisotropy and the CEF level scheme of TbPt$_6$Al$_3$, we performed INS experiments with an incident energy of 8.75 meV. A color-coded plot of INS intensity as a function of energy-momentum transfer $Q$ at 5 K is displayed in Fig. 8(a). At 5 K, three inelastic excitations are observed near 0.7, 1.9, and 3.8 meV, which are better seen in the $Q$-integrated (0 – 4 Å$^{-1}$) data in Fig. 8(b). The $J = 6$ multiplet of the Tb$^{3+}$ ion in a trigonal CEF is split into four non-Kramers doublets and five singlets. The trigonal point group $D_{3d}$ gives the CEF Hamiltonian for the single-ion model,

$$\mathcal{H}_{\text{CEF}} = B_2^0 O_2^0 + B_4^0 O_4^0 + B_4^3 O_4^3 + B_6^0 O_6^0 + B_6^3 O_6^3 + B_6^6 O_6^6,$$

where $B_n^m$ are CEF parameters and $O_n^m$ the Stevens operator equivalents [35]. Simultaneous fitting to the $B/M(T)$ and the INS data in the paramagnetic state using the software MANTID [30] gives the CEF parameters, $B_2^0 = 0.301 \times 10^{-1}$, $B_4^0 = 0.672 \times 10^{-3}$, $B_4^3 = 0.129 \times 10^{-1}$, $B_6^0 = 0.638 \times 10^{-5}$, $B_6^3 = -0.196 \times 10^{-3}$, and $B_6^6 = -0.879 \times 10^{-4}$ meV. The fitting result for $B/M(T)$ is displayed as $M(T)/B$ vs $T$ in Fig. 3. The fits to the INS data are represented in Fig 8(b), in which the energy differences from the ground state to the first, second, and third excited levels are obtained as 0.9, 1.9, and 4.1 meV. The wave function for the CEF ground state is described as $|\psi\rangle = \pm 0.295|\pm 5\rangle \mp 0.390|\pm 4\rangle -$

0.036|±2⟩ + 0.315|±1⟩ ± 0.496|∓1⟩ ± 0.023|∓2⟩ + 0.615|∓4⟩ + 0.187|∓5⟩ (double sign in same order). Since the magnetization component for the $a$ axis becomes zero in the CEF doublet ground state, the magnetic ordered moments of the $Tb^{3+}$ ion are not parallel to the $a$ axis but are along the $c$ axis. As shown in the inset of Fig. 5, the calculated curves of $M(B \parallel a)$ and $M(B \parallel c)$ at 10 K using the CEF parameters reproduce the data with the easy-plane magnetic anisotropy in the paramagnetic state. The calculation of $C_m$ is consistent with the Schottky anomaly at around 8 K shown in Fig. 6.

### E. Neutron powder diffraction

To determine the collinear AFM structure of $TbPt_6Al_3$ with the ordered moment along the $c$ axis, we carried out NPD experiments with a wavelength of 2.1969 Å. The Rietveld refinement to the NPD patterns was done using the software FULLPROF [36]. Figure 9 (a) represents the NPD pattern measured at 6 K > $T_N$. The Rietveld refinement of the crystal structure yielded the trigonal lattice parameters $a$ = 7.5052(1) Å and $c$ = 38.8017(4) Å. The NPD data were analyzed using a stoichiometric composition $TbPt_6Al_3$, together with an off-stoichiometric one $TbPt_{5.91}Al_{2.86}$. The Bragg $R$-factor for the latter is not smaller than that for the former. The weak peaks at around $2\theta$ = 57 and 67 deg are attributed to the impurity phase.

In the pattern at 0.7 K < $T_N$ (Fig. 9(b)), additional intensities appear on the top of the Bragg peak and other positions as marked by allows. All peaks can be indexed by the commensurate propagation vector **k** = [0, 0, 0], which has also been found to index the magnetic Bragg peaks in $NdPt_6Al_3$ [25] and $SmPt_6Al_3$ [26]. For the $R\bar{3}c$ space group with the magnetic propagation vector **k** = [0, 0, 0], the representational analysis using the BASIREPS [37] gives the irreducible representation for the Tb site as $\Gamma_{mag} = \Gamma_1 + \Gamma_2 + \Gamma_3 + \Gamma_4 + \Gamma_5 + \Gamma_6$. The basis vectors for each irreducible representations (IRs) are listed in Ref. [25]. $\Gamma_1$, $\Gamma_2$, $\Gamma_3$, and $\Gamma_4$ have only one basis vector along the $c$ axis. $\Gamma_3$ giving a ferromagnetic structure model with the parallel moments pointing along the $c$ axis is discarded. Since the direction of the ordered moment is expected to be parallel to the $c$ axis based on the CEF analysis, the four basis vectors along the $a$ and $b$ axes of $\Gamma_5$ and $\Gamma_6$ do not describe the collinear magnetic structure of $TbPt_6Al_3$. Out of the three IRs ($\Gamma_1$, $\Gamma_2$, and $\Gamma_4$), $\Gamma_1$ gives the best refinement of the diffraction pattern as shown in Fig. 9(b) with the smallest magnetic $R$-factor of 10.5. The magnetic structure model for $\Gamma_1$ is shown in Fig. 10(a), whose magnetic point group is $\bar{3}m.1$. The hexagonal two sublattices are coupled antiferromagnetically and two adjacent moments along the $c$ axis are stacked ferromagnetically. Figure

10(b) shows the temperature dependence of the refined magnetic moment. The blue line represents a fit in a temperature range between 2.5 and 3.5 K using an equation of $\alpha[(T_N - T)/T_N]^{2\beta}$, where $T_N = 3.7(1)$ K and the critical exponent $\beta = 0.29(6)$. The refined critical exponent $\beta \simeq 0.3$ is close to the expected value of 0.325 for 3D Ising model [38]. The size of the moment at 0.7 K is estimated as 5.1 $\mu_B$/ Tb, which is smaller than the value of 9.72 $\mu_B$ expected for the $Tb^{3+}$ free ion. This discrepancy is most likely due to the CEF effect [39].

### F. Altermagnetism

Finally, we discuss the active multipoles and altermagnetism in the magnetic point group of $\bar{3}m.1$ for $TbPt_6Al_3$. From a view of multipoles [4,12,40], the primary order parameter of the magnetic point group $\bar{3}m.1$ is expected as a magnetic toroidal monopole, magnetic toroidal quadrupole, and magnetic octupole. From a view of the altermagnetism [10,41], a nontrivial spin Laue group classifies the spin-momentum locking. The crystal structure of $TbPt_6Al_3$ has a $\bar{3}m$ crystallographic Laue group, and its subgroup of $\bar{3}$ interchanges the Tb atoms between the same-spin sublattices. The opposite-spin sublattices, on the other hand, are connected by the two-fold rotational symmetry on the honeycomb plane. Therefore, the nontrivial spin Laue group of $TbPt_6Al_3$ is identified as $^1\bar{3}^2m$, giving a $g$-wave altermagnet in which $CoF_3$, $FeF_3$, and $\alpha$-$Fe_2O_3$ have been classified [10,42,43]. In the $g$-wave altermagnet $TbPt_6Al_3$, the spin splitting of electron bands, chiral magnon, and piezomagnetic effect are expected. To observe such phenomena, single-crystal inelastic neutron scattering, angle-resolved photoemission spectroscopy, and magnetization measurements under uniaxial stress are highly anticipated.

### V. SUMMARY AND CONCLUSIONS

We studied the magnetic properties of $TbPt_6Al_3$ with a honeycomb structure of Tb ions. The transport, magnetic, and specific heat measurements in magnetic fields revealed a collinear AFM order at 3.5 K, where the $Tb^{3+}$ moments are pointing along the trigonal $c$ axis. The easy-plane magnetic anisotropy in the paramagnetic state, the Schottky anomaly, and the inelastic excitations are explained by the trigonal CEF model, in which the ground state is a non-Kramers doublet. Using the neutron powder-diffraction experiments, we determined the collinear AFM structure with the magnetic propagation vector $\mathbf{k} = [0, 0, 0]$. The nontrivial spin Laue group is determined as $^1\bar{3}^2m$, indicating that $TbPt_6Al_3$ is the first rare-earth-based *g*-wave altermagnet.

## ACKNOWLEDGMENTS

This work was supported by JSPS KAKENHI Grants No. JP21K03473, JP22J20278, JP22KJ2336, and JP23H04870. The low-temperature measurements and electron-probe microanalysis were performed at the Integrated Experimental Support/Research Division, N-BARD, Hiroshima University. We thank Y. Nambu and M. Ohkawara for their support for the NPD experiment. The NPD at HERMES, JRR-3 was carried out under the general user program managed by the Institute for Solid State Physics, University of Tokyo (Proposals No. 23615) and Institute for Materials Research, Tohoku University (Proposals No. 202212-CNKXX-0001). We acknowledge A. Kimura, H. Harima, S. Hayami, Y. Ogawa, T. Aoyama, H. Suzuki, K. Isobe, G. Balakrishnan, M. R. Lees, and D. A. Mayoh for fruitful discussions. D. T. Adroja would like to thank the Royal Society of London for the International Exchange funding between the U.K. and Japan and EPSRC-UK for grant No. EP/W00562X/1. We also thank ISIS for beam time on MARI (Exp No. RB2390042) [29].

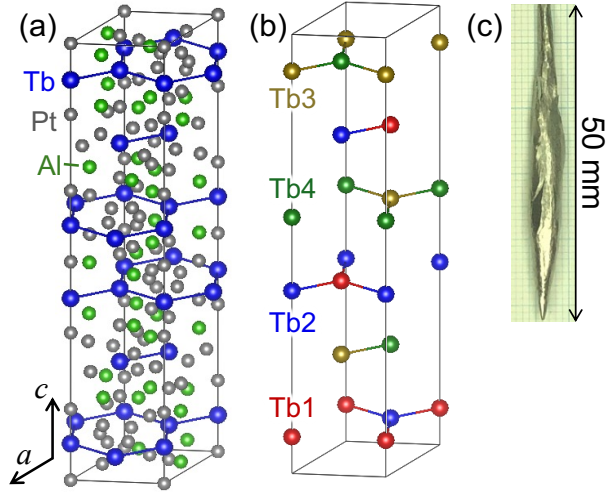

FIG. 1. (a) Crystal structure of TbPt$_6$Al$_3$ with the trigonal space group of $R\bar{3}c$. (b) There are four Tb atoms for Wycoff site 12$c$ in the unit cell: Tb1 at (0, 0, $z$), Tb2 (0, 0, $-z + 0.5$), Tb3 (0, 0, $-z$), and Tb4 (0, 0, $z + 0.5$), where $z = 0.084$. (c) Photo of the single-crystalline sample of TbPt$_6$Al$_3$ grown by the Czochralski method.

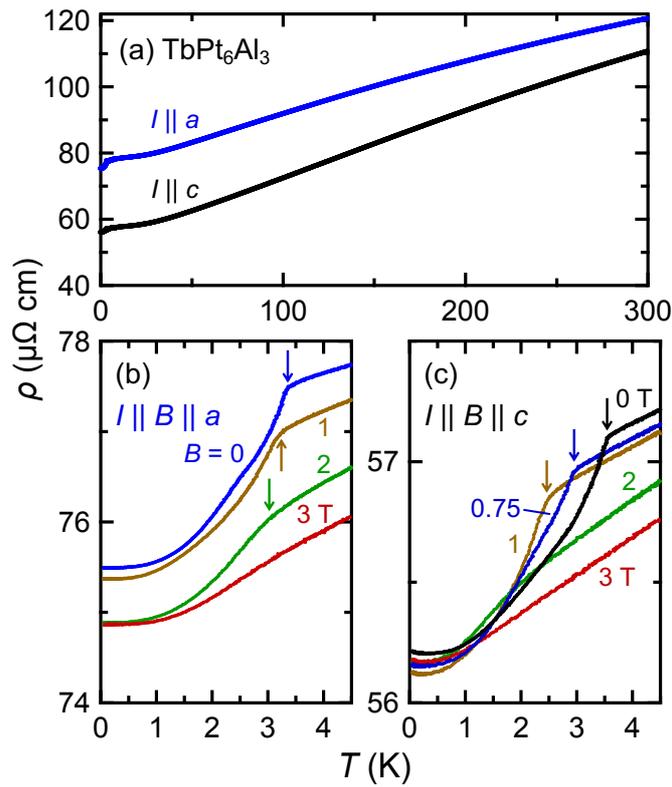

FIG. 2. (a) Temperature dependences of the electrical resistivity $\rho(T)$ of TbPt$_6$Al$_3$ for the current directions $I \parallel a$ and $I \parallel c$. Low-temperature $\rho(T)$ data in longitudinal magnetic fields of (b) $B \parallel a$ and (c) $B \parallel c$.

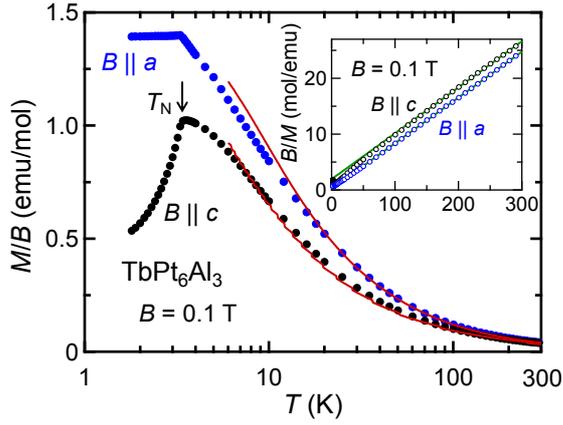

FIG. 3. Temperature dependences of the magnetic susceptibility $M(T)/B$ of TbPt$_6$Al$_3$ for the magnetic fields $B \parallel a$ and $B \parallel c$. The solid lines in red represent the calculated $M(T)/B$ data by using the CEF model for the Tb$^{3+}$ ion (see text). The inset shows the inverse of the $M(T)/B$ data. The solid lines in green are fits with a Curie-Weiss form to the data for $T > 100$ K.

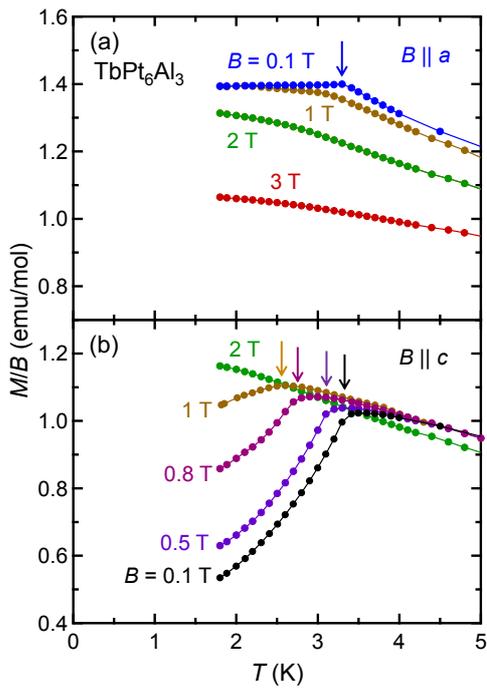

FIG. 4. Temperature dependences of $M(T)/B$ data of TbPt$_6$Al$_3$ in different magnetic fields (a) $B \parallel a$ and (b) $B \parallel c$.

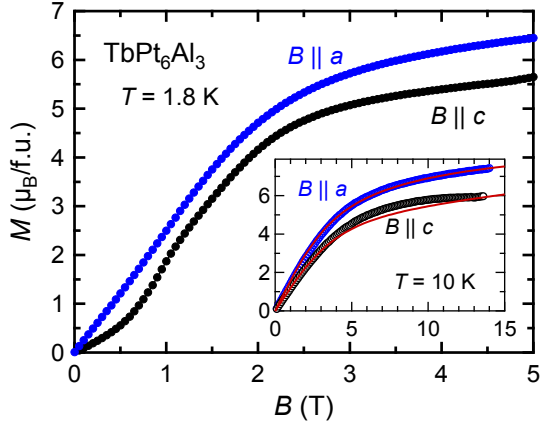

FIG. 5. Isothermal magnetization $M(B)$ of TbPt$_6$Al$_3$ at 1.8 K for $B \parallel a$ and $B \parallel c$. The inset shows the $M(B)$ data at 10 K $>T_N$ and the calculated ones (solid lines in red) using the CEF model for the Tb$^{3+}$ ion.

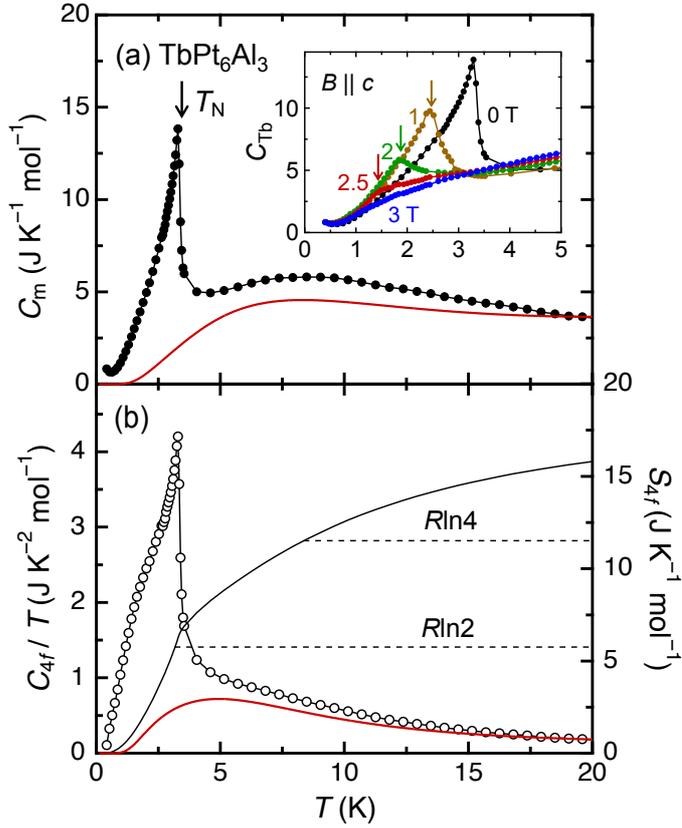

FIG. 6. (a) Temperature dependence of the magnetic specific heat data, $C_m = C_{Tb} - C_Y$, which were calculated by subtracting the $C(T)$ for YPt$_6$Al$_3$ from that for TbPt$_6$Al$_3$. The inset shows the $C_{Tb}$ data in constant magnetic fields applied parallel to the $c$ axis. (b) Temperature dependences of 4$f$ electron contribution to the specific heat $C_{4f}$ divided by temperature (left-hand scale) and its entropy $S_{4f}$ (right-hand scale). The $C_{4f}$ data were estimated by subtracting the nuclear Schottky contribution from $C_m$ (see text). The solid lines (red) in (a) and (b) are calculated by the CEF model for the Tb$^{3+}$ ion (see text).

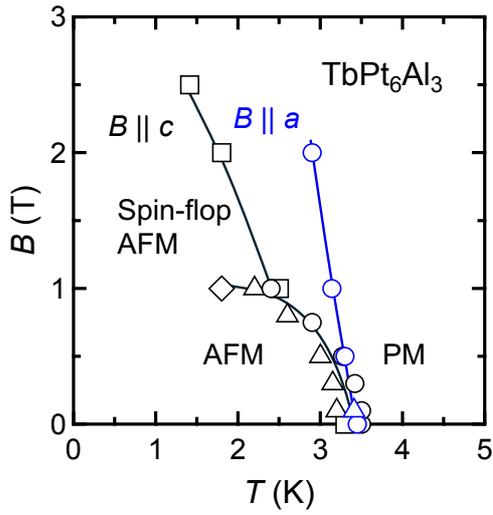

FIG. 7. Magnetic fields vs temperature phase diagram of TbPt$_6$Al$_3$ for $B \parallel a$ and $B \parallel c$ constructed from anomalies in magnetic susceptibility (triangles), isothermal magnetization (diamond), electrical resistivity (circles), and specific heat (squares).

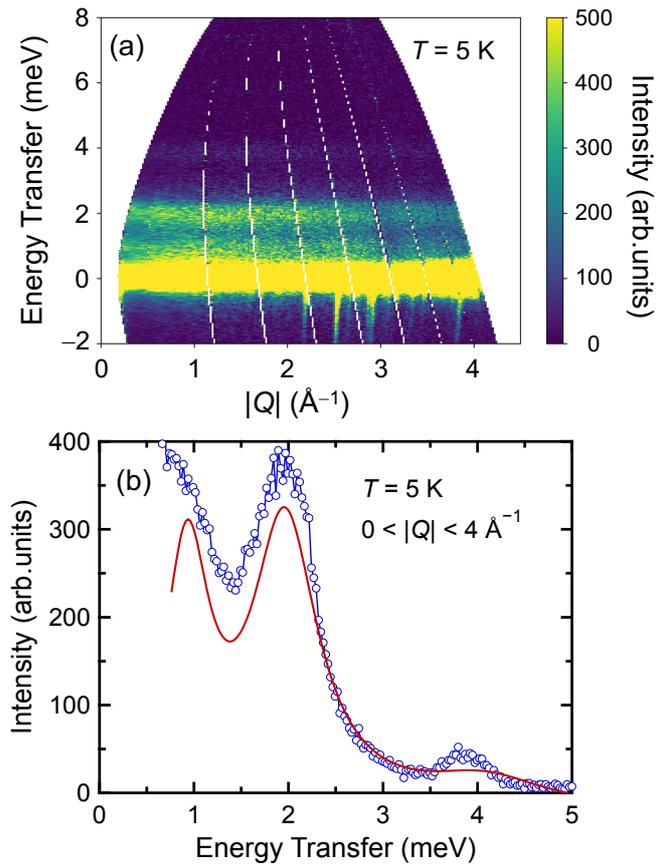

FIG. 8. (a) Color-coded plots of the INS intensity for TbPt$_6$Al$_3$ at 5 K with an incident energy of $E_i$ = 8.75 meV. (b) Intensity as a function of energy transfer. The solid line in red shows the fit using the trigonal CEF model for the Tb$^{3+}$ ion.

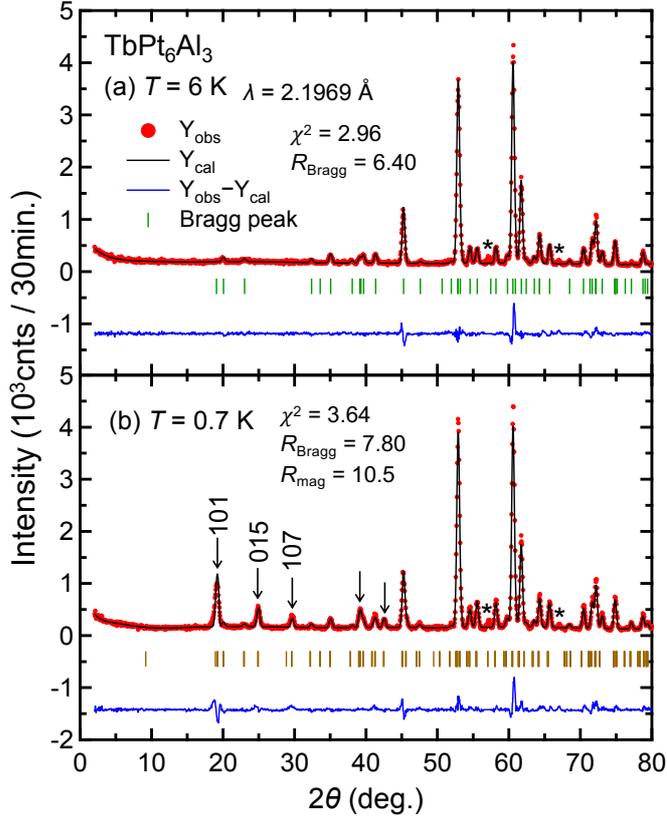

FIG. 9. Rietveld refinement of the neutron powder diffraction patterns of TbPt$_6$Al$_3$ at (a) 6 K and (b) 0.7 K. The red points are the experimental data, the black line is the calculated pattern, and the blue line is the difference curve. The green and brown vertical ticks represent nuclear Bragg and magnetic Bragg peak positions, respectively.

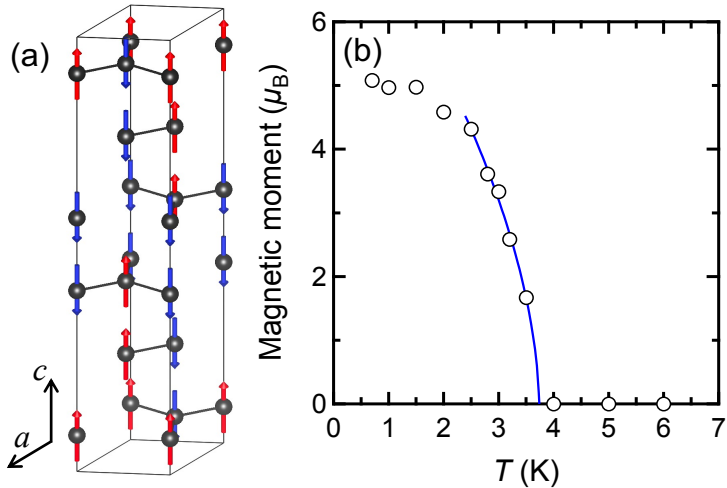

FIG. 10. (a) Magnetic structure of TbPt$_6$Al$_3$ for the magnetic point group $\bar{3}m.1$ of TbPt$_6$Al$_3$. (b) Temperature dependence of the magnetic moments of the Tb$^{3+}$ ions. The solid line in blue shows a fit with a form of $\alpha[(T_N - T)/T_N]^{2\beta}$ (see text).